\documentclass{aa}

\usepackage{graphicx}
\usepackage{amsmath}
\usepackage{xspace}
\usepackage{txfonts}
\usepackage{hyperref}
\newcommand\corot[1][]{CoRoT-7#1\xspace}
\newcommand\given[1][]{\:#1\vert\:}
\newcommand{\rhk}{$\log R'_{\rm HK}$\xspace}
\DeclareMathOperator{\ms}{\,m\,s^{-1}} 

\begin{document}

   \title{Uncovering the planets and stellar activity of \corot \\ using only radial velocities%
          \thanks{All data and software presented in this article are available online at \url{https://github.com/j-faria/exoBD-CoRoT7}.}
   }
   \titlerunning{Uncovering the planets and stellar activity of \corot using only radial velocities}

   \author{%
    J. P. Faria\inst{1,2}
    \and
    R. D. Haywood\inst{3}
    \and
    B. J. Brewer\inst{4}
    \and
    P. Figueira\inst{1}
    \and
    M. Oshagh\inst{1,5}
    \and
    A. Santerne\inst{1}
    \and
    N. C. Santos\inst{1,2}
   }
   
   \authorrunning{J. P. Faria et al.}

   \institute{%
      Instituto de Astrofísica e Ciências do Espaço, Universidade do Porto, CAUP, Rua das Estrelas, PT4150-762 Porto, Portugal\\
      \email{joao.faria@astro.up.pt}
      \and
      Departamento de Física e Astronomia, Faculdade de Ciências, Universidade do Porto, Rua Campo Alegre, 4169-007 Porto, Portugal
      \and
      Harvard-Smithsonian Center for Astrophysics, 60 Garden Street, Cambridge, MA 02138, USA
      \and
      Department of Statistics, The University of Auckland, Private Bag 92019, Auckland 1142, New Zealand
      \and
      Institut f\"ur Astrophysik, Georg-August-Universit\"at, Friedrich-Hund-Platz 1, 37077 G\"ottingen, Germany
    }

   \date{Received December 7, 2015; accepted January 27, 2016}

 
  \abstract
  {%
  Stellar activity can induce signals in the radial velocities of stars, complicating the detection of orbiting low-mass planets.
  We present a method to determine the number of planetary signals present in radial-velocity datasets of active stars, using only radial-velocity observations.
  Instead of considering separate fits with different number of planets, we use a birth-death Markov chain Monte Carlo algorithm to infer the posterior distribution for the number of planets in a single run. 
  In a natural way, the marginal distributions for the orbital parameters of all planets are also inferred.
  This method is applied to HARPS data of \corot.
  We confidently recover the orbits of both \corot[b] and \corot[c] although the data show evidence for the presence of additional signals.
  }


   \maketitle
%


\section{Introduction}

  Imagine we have at our disposal a set of spectroscopic observations of an unknown star, and we can obtain precise radial-velocity (RV) measurements, using the cross-correlation function (CCF) technique \citep{Baranne1996,Pepe2002}.
  The so-called line profile indicators (such as the full width at half maximum of the CCF, its bisector span, and associated quantities, e.g. \citealt{Figueira2013}) and activity indicators, like the \rhk \citep{Noyes1984}, are also usually available but are not necessary for our analysis.
  And without other instruments at hand, we cannot measure the star's photometric variations, as is the case for most targets in RV surveys.

  Armed with some statistical artillery, we aim to answer the following question: how many orbiting planets can be confidently detected in these data, and what are their orbital parameters and minimum masses?

  Besides planets, other physical processes can induce variations in the radial velocities of a star.  
  These include stellar oscillations, granulation, spots and faculae/plages, and long-term magnetic activity cycles \citep[see, e.g.][]{Saar1997,Santos2010a,Boisse2011,Dumusque2011a}.
  Some of these activity-induced signals can be mitigated or averaged out by adapting the observational strategy \citep{Dumusque2011b}.
  But signals caused by the presence of active regions  on the stellar surface can show periodicities and amplitudes similar to the ones induced by real planetary signals, and may be harder to disentangle. 
  Indeed, these signals can even mimic planetary signals \citep[e.g.][]{Figueira2010,Santos2014,Robertson2015a}.
  
  Simultaneous photometric and RV observations have  been successfully used to constrain activity-induced RV signals. 
  However, this approach requires either a joint model for photometric and RV variations, which can be statistical \citep[e.g.][]{Rajpaul2015} or based on a description of the stellar features inducing the signal \citep[e.g.][]{Lanza2010}, or a conversion from photometric variations to RV variations \citep[e.g.][hereafter H14]{Haywood2014}.
  \defcitealias{Haywood2014}{H14}

  Here we present a framework that models activity-induced signals as correlated noise in the RV observations and does not require simultaneous photometric observations.
  The method is based on the fact that, for data of sufficient quality, it is possible to distinguish if an oscillating signal has the Keplerian shape that is expected from a planet, or some other approximately-periodic shape as expected from stellar activity.
  Below we describe our model and apply it to High Accuracy Radial Velocity Planet Searcher (HARPS; \citealt{Mayor2003}) observations of \corot.



\section{A Bayesian model for RV data}\label{sec:RVmodel}

  In this section we present in detail our model for RV data and describe the two types of signals we consider: planetary signals and activity-induced noise.

  \subsection{RV planetary signals}

  The dynamical evolution of a planetary system is governed by the gravitational interactions of its constituent bodies. 
  For most systems with multiple planets one can assume, to a good approximation, that the mutual planetary perturbations are negligible on timescales that are comparable to the duration of observations. 
  The stellar RV perturbations that are due to a multiple-planet system can then be modelled as a linear superposition of Keplerian orbits.

  Each Keplerian can be described with five RV observables: the semi-amplitude $K$, the orbital period $P$, the eccentricity $e$, an orbital phase $\phi$, and the longitude of the line-of-sight, $\omega$. 
  We also consider a systemic velocity, $v_{\rm sys}$, of the centre of mass of the system, which corresponds to an RV zero-point measured by HARPS.

  \subsection{Gaussian processes to model stellar activity}

  Even if activity signals cannot be easily described analytically (a complete description would require knowledge of the active region distribution and evolution, temperature contrast and stellar parameters, such as inclination and limb darkening), we can make some assumptions about their form. 
  Besides assuming the signals will be continuous and smooth, stellar rotation  induces a periodicity or quasi-periodicity, as active regions evolve and cycle in and out of view on the stellar surface.

  For the purposes of detecting planets using RV measurements, the signals caused by stellar activity can then be seen as correlated quasi-periodic noise. 
  Gaussian processes (GP) are an increasingly common tool to deal with correlated noise \citep[e.g.][]{Roberts2012}. 
  In their application to regression problems, GPs can be seen as prior distributions over functions, which will be constrained by the data \citep[e.g.][]{Rasmussen2006}.
  For our purposes, we use the GP to model the stochastic component of our signal -- that is, the stellar noise. 

  A GP is defined by its mean function, the deterministic component of the signal, and its covariance function, which defines the overall behaviour of the functions under the GP distribution. 
  When the covariance function is evaluated at the observed times,  the covariance matrix is obtained.
  From the many possible choices for a covariance function, the quasi-periodic kernel is the most widely used in the exoplanet literature (e.g. \citetalias{Haywood2014}, \citealt{Grunblatt2015,Rajpaul2015}), which results in a covariance matrix of the form
  \begin{equation}\label{eq:covariance_matrix}
    \Sigma_{ij} = \eta_1^2 
                    \exp\left[ - \frac{(t_i - t_j)^2}{2\eta_2^2} 
                              - \frac{2\sin^2\left(\frac{\pi (t_i-t_j)}{\eta_3}\right)}{\eta_4^2} \right] 
                    + \left( \sigma_i^2 + s^2 \right) \delta_{ij} \:\text{.}
  \end{equation}
  
  This represents some of our expectations for the activity-induced RV signals: the correlations decay on a timescale of $\eta_2$ days and \textbf{[Note 1: the suggested 'to' changes the meaning of the sentence.]} have a periodic component with period $\eta_3$ days.
  The parameter $\eta_4$ controls the relative importance of the periodic and decaying components.
  The parameter $\eta_1$ represents the amplitude of the correlations. 
  This covariance matrix also takes additional uncorrelated noise into account, added quadratically to the diagonal, where $\sigma_i$ are the reported RV uncertainties and $s$ is a free parameter.

  \subsection{The complete model}

    \begin{figure}
    \centering
    \includegraphics[width=\hsize]{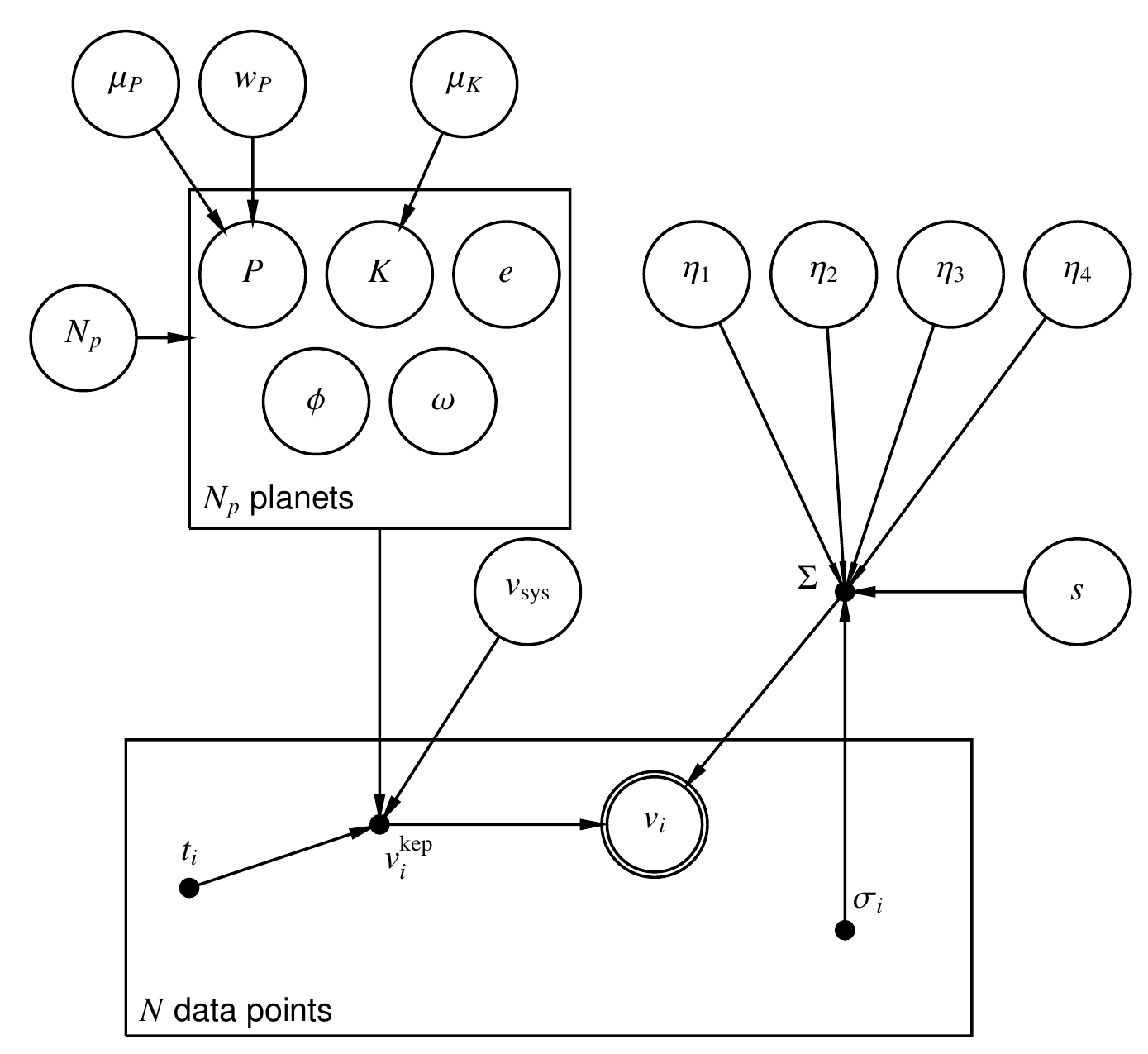}
    \caption{Representation of the relations between parameters and observations in our 
             RV model, as a probabilistic graphical model. 
             An arrow between two nodes indicates the direction of conditional dependence.
             The circled nodes are the parameters of the model, whose joint distribution is sampled by the Markov chain Monte Carlo (MCMC) algorithm.
             The double circled node $v_i$ represents the observed RVs.
             The filled nodes represent deterministic variables: if these variables have parent nodes ($v_i^{_{\rm kep}}$ and $\Sigma$), they are given by a deterministic function of those parents; if they do not have parents ($t_i$ and $\sigma_i$), they are assumed as being given and thus fixed.
             The variables inside boxes are repeated a number of times, as shown in the bottom left corner of each box.}
    \label{fig:PGM}%
    \end{figure}

  Our complete model for the RV observations is shown in Fig. \ref{fig:PGM}, in the form of a probabilistic graphical model. 
  The diagram relates all the parameters (see Table \ref{tab:priors} for their description) and observables in the model, stating their inter-dependencies.
  From this graphical representation we can build an expression for the joint probability of the parameters of interest and the data.

  Let $\Theta$ be the vector of all the parameters in the model: 
  $
    \Theta = \left[ N_p, \mu_P, w_P, \mu_K, \{P, K, e, \phi, \omega\}, v_{\rm sys}, \eta_1, \eta_2, \eta_3, \eta_4, s \right]
  $.
  The notation means that the subset $\{P, K, e, \phi, \omega\}$ will be repeated $N_p$ times (for each of the $N_p$ planets). 
  An observed dataset is composed of $N$ radial-velocity observations $v_i$, at times $t_i$ and with associated uncertainty estimates $\sigma_i$. 
  The diagram in Fig. \ref{fig:PGM} is a representation of the joint probability density function (PDF) for all the variables:
  \begin{equation*}
    p\left( \Theta, \{v_i\}, \big\{v_i^{_{\rm kep}}\big\} \given[\big] \{t_i\}, \{\sigma_i\}, \mathcal{I} \right) \text{,}
  \end{equation*}
  where we condition on the information $\mathcal{I}$\footnote{Here, $\mathcal{I}$ encodes the assumptions considered when setting up the problem, e.g. the form of the GP kernel or the fact that we ignore planet-planet interactions, etc.}.
  In the remainder of the paper we will include $\{t_i\}$ and $\{\sigma_i\}$ in $\mathcal{I}$, because they are assumed to be fixed.
  To ease the notation, we also group the RV values in a proposition $D=\{v_i\}$.
  The $\big\{v_i^{_{\rm kep}}\big\}$ are obtained from evaluating the sum of Keplerian signals at the observed times.

  The joint PDF can be factored and rearranged to give Bayes's theorem,
  \begin{equation}\label{eq:bayes}
    p(\Theta \given D,\mathcal{I}) = \frac{p (\Theta \given \mathcal{I} ) \: p(D \given \Theta, \mathcal{I})}{p(D \given \mathcal{I})} \,,
  \end{equation}
  providing an expression for the posterior distribution for all the parameters, conditioned on the observed data.
  The posterior distribution contains all the information about the parameters that is available to us. 
  With this distribution to hand, we can then infer how many planets are supported by the data and their orbital parameters and masses.

  \subsection{Determining the number of planets}

    To calculate the joint posterior distribution, following Eq. \eqref{eq:bayes}, we need the three terms on the right-hand side: the prior $p (\Theta \given \mathcal{I} )$, likelihood $p(D \given \Theta, \mathcal{I}),$ and evidence $p(D \given \mathcal{I})$. 

    For the likelihood, the choice that reflects most genuinely our state of knowledge is a multivariate Gaussian distribution\footnote{Fig. \ref{fig:PGM} shows that all we know about the distribution of the $\{v_i\}$ are its first and second moments. The multivariate Gaussian follows from the principle of maximum entropy when the distribution is constrained to having a specified covariance matrix \cite[e.g.][]{Cover2006}.}, with mean $\big\{v_i^{_{\rm kep}}\big\}$ and covariance matrix $\Sigma$. 
    The complete covariance matrix can be obtained, deterministically, from the values of $\eta_1, \eta_2, \eta_3, \eta_4, \sigma_i$, and $s$, according to Eq. \eqref{eq:covariance_matrix}.
    The log-likelihood is then given by:
    \begin{align}
      \ln p(D \given \Theta, \mathcal{I}) = -\frac{1}{2} \mathbf{r}^T \, \Sigma^{-1} \, \mathbf{r} 
                                  - \frac{1}{2} \ln \det \Sigma
                                  - \frac{N}{2} \ln 2\pi \text{,}
    \end{align}
    where $\mathbf{r}$ is the vector given by $v_i - v_i^{_{\rm kep}}$ for all data points.

    The priors used for all the parameters are listed in Table \ref{tab:priors}.
    Most of them are the same as those used by \citet{Brewer2015} and were chosen to represent uninformative or vague prior knowledge.
    For the orbital periods and semi-amplitudes we assign hierarchical priors that are conditional on the hyperparameters, $\mu_P$, $w_P$, and $\mu_K$, respectively (see Table \ref{tab:priors}).
    This reflects our belief that knowing the parameters of one planet provides a small amount of information about the parameters of another planet.

      \begin{table}[h]
      \caption{Meaning and prior distribution for the parameters in the model. In some cases we sample on the logarithm of the parameter.}
      \label{tab:priors}
      \centering    
      \begin{tabular}{clcc}
      \hline\hline

      \multicolumn{4}{c}{hyper parameters}  \\
      \hline
      $N_p$      & number of planets         &        & $\mathcal{U}\,(0, 10)$       \\ 
      $\mu_P$    & median orbital period     & $\log$ & $\mathcal{C}\,(5.9, 1)$               \\
      $w_P$      & diversity orbital periods &        & $\mathcal{U}\,(0.1, 3)$               \\
      $\mu_K$    & mean semi-amplitude       & $\log$ & $\mathcal{C}\,(0, 1)$                 \\[0.8ex]
      \hline           
      \multicolumn{4}{c}{planet parameters}  \\
      \hline 
      $P$        & orbital period                 & $\log$ & $\mathcal{L}\,(\log \mu_P, w_P)$ \\
      $K$        & semi-amplitude                 &        & $\mathcal{E}\,(\mu_K)$           \\
      $e$        & eccentricity                   &        & $\mathcal{B}\,(1, 3.1)$     \\
      $\phi$     & orbital phase                  &        & $\mathcal{U}\,(0, 2\pi)$         \\
      $\omega$   & longitude of line-of-sight     &        & $\mathcal{U}\,(0, 2\pi)$                \\[1ex]
      \hline
      \multicolumn{4}{c}{GP and noise parameters}  \\
      \hline           
      $\eta_1$ & amplitude of covariance &        & $\mathcal{LU}\,(0.1, 50)$  \\
      $\eta_2$ & aperiodic timescale     &        & $\mathcal{LU}\,(1, 100)$   \\
      $\eta_3$ & correlation period      &        & $\mathcal{U}\,(10, 40) $   \\
      $\eta_4$ & periodic scale      &        & $\mathcal{LU}\,(0.1, 10)$  \\
      $s$ & extra white noise            & $\log$ & $\mathcal{C}(0, 1)$        \\[1ex]
      \hline
      $v_{\rm sys}$ & systematic velocity && $\mathcal{U}\,(\min v_i, \max v_i)$ \\
      \end{tabular}
      \tablefoot{Symbol meaning: 
                 $\mathcal{U}(\cdot, \cdot)$ -- Uniform prior with lower and upper limits;
                 $\mathcal{C}(\cdot, \cdot)$ -- Cauchy prior with location and scale (these distributions were truncated for numerical reasons);
                 $\mathcal{L}(\cdot, \cdot)$ -- Laplace prior (sometimes called double exponential) with location and scale;
                 $\mathcal{E}(\cdot)$ -- Exponential prior with mean;
                 $\mathcal{B}(\alpha, \beta)$ -- Beta prior with shape parameters, $\alpha$ and $\beta$, an approximation to the frequency distribution of eccentricities proposed by \citet{Kipping2013};
                 $\mathcal{LU}(\cdot, \cdot)$ -- Log-uniform prior with lower and upper limits.}
      \end{table}

    For most of the parameters of the GP covariance kernel, we assigned log-uniform priors in sensible ranges.
    For $\eta_3$, the parameter that can be interpreted as the stellar rotation period, we assumed a uniform prior between \textbf{[Note 2: I believe 'between' is better suited here than the suggested 'of']} 10 and 40 days.

    To sample the joint posterior distribution, we used the algorithm proposed by \citet{Brewer2014}. 
    The particular difficulty here -- since $N_p$ is not known -- is that the sampling algorithm needs to jump between candidate solutions with different numbers of planets. 
    \citet{Brewer2014} proposed a method that uses birth-death Markov chain Monte Carlo (MCMC) moves to infer $N_p$, within the diffusive nested sampling framework \citep{Brewer2011}. 
    This way, one can estimate the value of the evidence while improving the mixing in complex posteriors (affected by multimodality and phase transitions).
    \citet{Brewer2015} applied this method to RV data of $\nu$ Oph and Gliese 581, although these authors did not incorporate GPs in their analysis.

\section{Application to HARPS data} 
\label{sec:application_to_real_data}

  We apply the method described in the previous section to HARPS observations of \corot.
  The planet \corot[b] was first announced by \citet{Leger2009} and was the first super-Earth with a measured radius.
  Its orbital period is estimated from the transits in the CoRoT light curve as being $P_{\rm b} = 0.854$ days \citep{Leger2009}. 
  A second non-transiting planet, with $P_c = 3.69$ days, was detected in a follow-up RV campaign \citep{Queloz2009} and a more disputed detection of a third planetary signal was reported by \citet{Hatzes2010}.

  Owing to the high activity levels of the host star, this system has since generated a wealth of discussion, resulting in different estimates for the masses of the planets \citep{Lanza2010,Boisse2011,Ferraz-Mello2011,Pont2011a,Hatzes2011}.
  Simultaneous observations from CoRoT and HARPS were obtained in 2012 to help settle these issues (\citealt{Barros2014}; \citetalias{Haywood2014}).
  These observations were analysed by \citetalias{Haywood2014} with an RV model that is similar to ours but considering information from the simultaneous photometric observations.
  
  We emphasise that, in the following, we analyse the full set of RV observations.
  In summary, the star was observed with HARPS in 2009 and 2012, with a total of 177 public RV measurements.
  The average error bar on these measurements is $2\ms$ (which includes photon and instrumental noise) and the RV dispersion is $10\ms$, over the complete 3-year timespan.

  A common procedure in (current) RV studies is to fix the number of planets and sample the posterior for the remaining parameters (and possibly calculate the evidence) in a step-by-step approach.
  If we fix $N_p = 2$, we recover the orbital parameters of the two known planets, \corot[b] and \corot[c], within the errors reported in \citetalias{Haywood2014} and \citet{Barros2014}.
  The method we present in this paper is, however, much more general and allows for the full posterior for $N_p$ to be obtained from one run.
  We now proceed with this general method.

\begin{figure}
    \centering
    \includegraphics[width=\hsize]{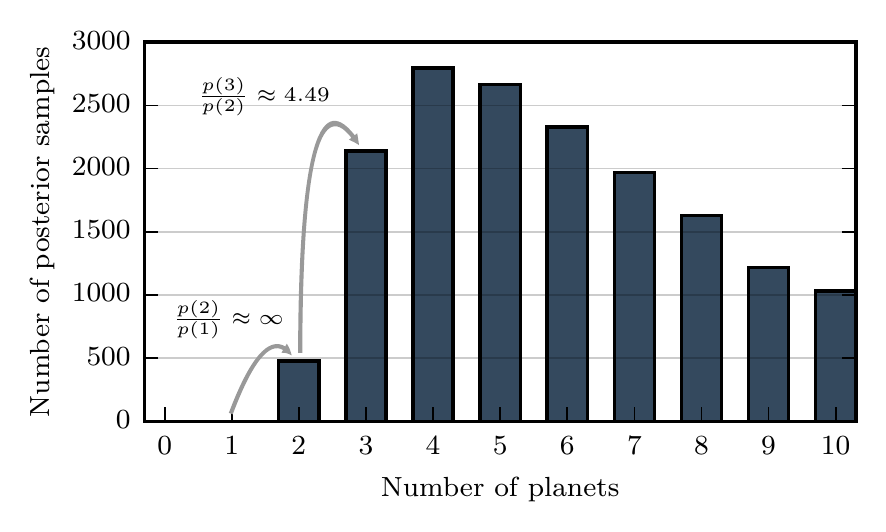}
    \caption{Posterior distribution for the number of planets $N_p$.
             The counts are number of posterior samples in models with a given number of planets.
             The two ratios of probabilities between models with 1, 2, and 3 planets is highlighted; note that $p(0) = p(1) = 0$.}
    \label{fig:posteriorNp}
  \end{figure}

  We ran our algorithm on the full set of RVs, using the priors in Table \ref{tab:priors}, and obtained 16\,248 effective samples from the joint posterior distribution\footnote{%
    The computer used to run the simulations was equipped with an Intel\textsuperscript{\textregistered} Core\textsuperscript{TM} i5-4460 CPU running at 3.20 GHz and 4 GB of RAM.
    The running time to obtain 50 000 samples (from diffusive nested sampling's target mixture distribution) was four days.
    Since the computational cost of the algorithm is dominated by the inversion of the covariance matrix, it is expected to scale roughly with $N^3$.}.
  The evidence for our model is $\log(p(D \given \mathcal{I})) = -530.9$.
  The resulting marginal posterior distribution for $N_p$ is shown in Fig. \ref{fig:posteriorNp}.

    \begin{figure}
    \centering
    \includegraphics[width=\hsize]{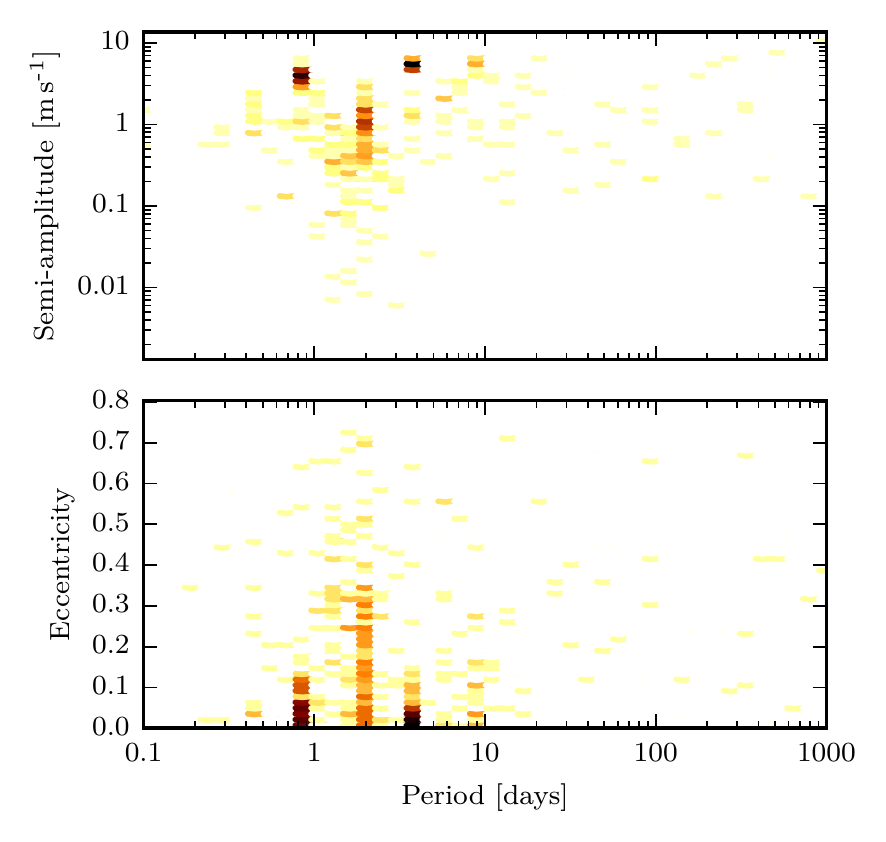}
    \caption{Joint posterior distribution for the semi-amplitudes (top panel) and the eccentricities (bottom panel) together with the orbital periods of the Keplerian signals.}
    \label{fig:jointposterior}%
    \end{figure}

  The posterior distribution for $N_p$ is one of the main outputs of our method. 
  But  to actually decide on what is the number of planets orbiting \corot, and answer our initial question, we need to clarify what we mean by ``confidently detected''. 
  Based on a scale suggested by \citet{Jeffreys1998} (see also \citealt{Kass1995}), some authors (e.g. \citealt{Tuomi2011,Feroz2011}) have proposed that, to claim a detection of $N_p$ planets, the probability of $N_p$ should be 150 times greater than the probability of $N_p - 1$.
  This criterion requires 'strong' \citep{Jeffreys1998} evidence for detecting a planet, thus considering
  false positives to be much worse than false negatives.
  By applying this rule to our results, we choose $N_p = 2$ as the number of planets confidently detected in our dataset.

  We emphasise that having confidently detected two planets is a different matter from knowing that the number of planets is actually two. 
  According to the posterior distribution, it is more likely that there are four planets. 
  But our detection criterion depends on the ratio of probabilities for consecutive values of $N_p$ (see Fig. \ref{fig:posteriorNp}), not on the probability values themselves.
  Of course, the posterior distribution is sensitive to the prior distributions, in particular as to how small we believe $K$ might be.

  The joint posterior distributions for the orbital periods, semi-amplitudes, and eccentricities of the signals are shown in Fig. \ref{fig:jointposterior}, where the samples for all Keplerians were combined.
  The figure shows a 2-dimensional histogram of the posterior samples, where the colourmap represents bin counts and is set in a logarithmic scale.

  The two detected planets are seen as overdensity regions at $P_{\rm b}=0.85$ days and $P_{\rm c}=3.69$ days.
  Their amplitudes and eccentricities are well constrained.
  There is a clear posterior peak around two days with amplitude and eccentricity mostly unconstrained.
  The posterior also shows a smaller peak at 9 days, the period reported by \citet{Hatzes2010} as a possible third planet \citep[see also][]{Tuomi2014}.
  
\begin{figure*}
  \centering
  \includegraphics[width=\hsize]{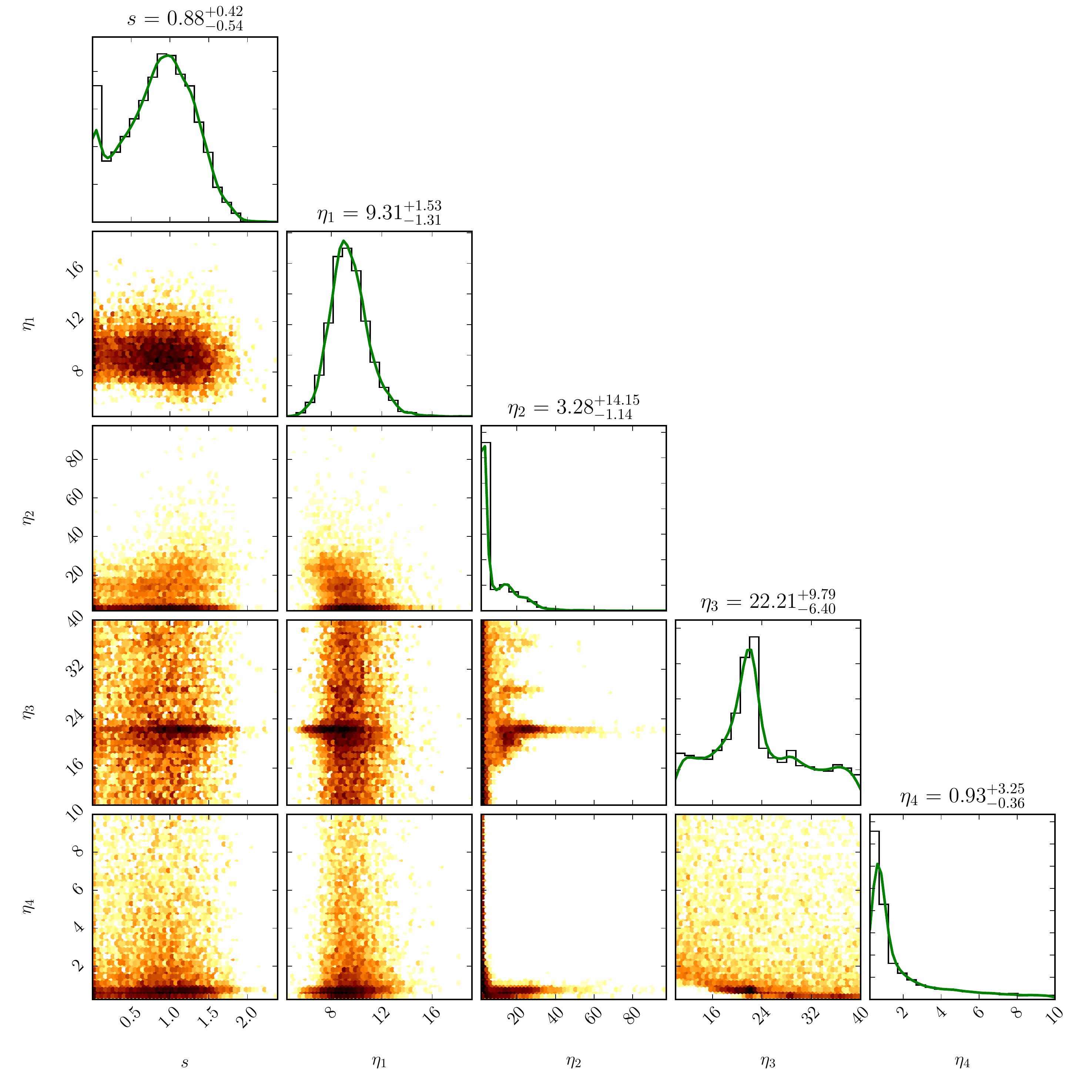}
  \caption{Marginalised 1- and 2-D posterior distributions for the parameters of the GP and the extra white noise. The samples for all values of $N_p$ were combined. The titles above each column show the median of the posterior and the uncertainties calculated from the 16\% and 84\% quantiles. The solid lines are kernel density estimations of the marginal distributions.}
  \label{fig:cornerGP}
  \end{figure*}

  Marginal posterior distributions for the parameters of the GP and the extra white noise parameter $s$ are shown in Fig. \ref{fig:cornerGP}. 
  The posterior for $\eta_3$ is particularly interesting as it provides a constraint to \corot's rotation period, obtained exclusively from the RVs.
  Our inferred value for the stellar rotation period of $22.30^{+10.08}_{-6.11}$ days is obtained solely from the RV time-series and is in agreement (see Table \ref{tab:parameters}) with earlier estimates which used the CoRoT light curve (\citealt{Leger2009,Lanza2010}; \citetalias{Haywood2014}).

  Also interesting is the joint behaviour of $\eta_2$, $\eta_3$, and $\eta_4$.
  For higher values of $\eta_4$, the periodic component of the covariance function loses importance relative to the decaying component, $\eta_2$ gets smaller and $\eta_3$ becomes unconstrained.
  In this case, the GP smooths the RVs on a timescale of $\eta_2 \approx 3$ days.
  But when $\eta_4$ is of order unity (meaning the periodic component is present), the decaying timescale is higher and $\eta_3$ is constrained around 22 days.
  The values of $\eta_2$ in this situation (20-30 days) are closer to the stellar rotation period and are also consistent with the average lifetime of active regions measured in the CoRoT 2012 photometry \citepalias[20.6 $\pm$ 2.5 days, ][]{Haywood2014}.
  Our results therefore validate the approach taken by, e.g. \citetalias{Haywood2014} and \citet{Grunblatt2015}, of modelling the activity-induced RV variations with a GP that has the covariance properties of the light curve.

\begin{figure}
    \centering
    \includegraphics[width=\hsize]{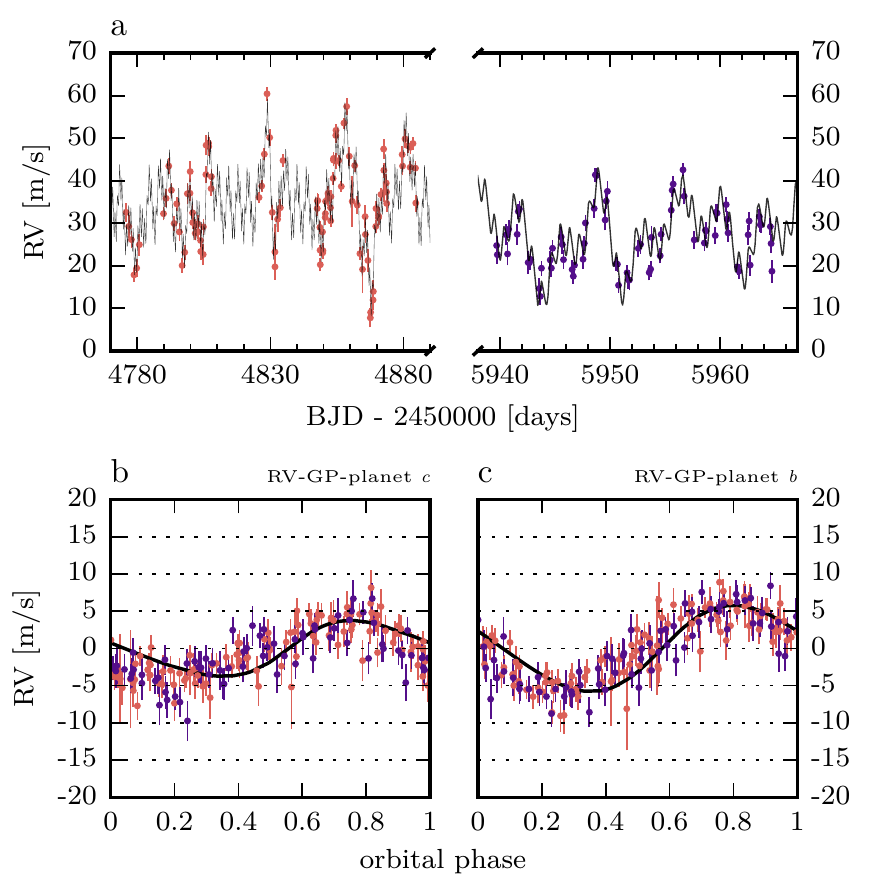}
    \caption{Panel a: RV measurements of \corot from 2009 and 2012 and the two-planet best-fit model (black curve).
             Note the different scales in the abscissae on the left and right parts of the plot.
             Panels b and c show the phased RV curves after subtracting each planet signal and the GP.}
    \label{fig:2planet-fit}%
    \end{figure}

  Considering only the posterior samples with $N_p = 2$, Table \ref{tab:parameters} lists the median values of some orbital parameters for the two planets, and the maximum likelihood RV curves are shown in Fig. \ref{fig:2planet-fit}.
  Our estimates for the orbital parameters are in agreement, within the uncertainties, with the ones obtained by \citetalias{Haywood2014}.

      \newcommand\Tstrut{\rule{0pt}{2.6ex}}         
      \newcommand\Bstrut{\rule[-0.9ex]{0pt}{0pt}}   
      \newcommand\givespace{\Tstrut\Bstrut\\}

      \begin{table}
      \caption{Parameter estimates from our work and from \citetalias{Haywood2014}. We consider all models that have $N_p = 2$ and show the marginal posterior medians, together with the 16\% and 84\% quantiles.
      }
      \label{tab:parameters}
      \centering    
      \begin{tabular}{lccc}
      \hline\hline

      ~ & units & This work & \citetalias{Haywood2014}  \givespace 
      \hline
      $P_{\rm{b}}$  \tablefootmark{~(a)} & days     & $0.85424\,^{+\,0.00071}_{-\,0.00126}$ & $0.85359165 \pm \, 5\!\times\!10^{-8}$ \givespace
      $K_{\rm{b}}$  & $\ms$     & $3.97\,^{+\,0.62}_{-\,0.55}$          & $3.42 \pm 0.66$ \givespace
      $e_{\rm{b}}$  &           & $0.045\,^{+\,0.053}_{-\,0.027}$       & $0.12 \pm 0.07$ \givespace
      $m_{\rm{b}}$  & $M_\oplus$  & $5.53\,^{+\,0.86}_{-\,0.78}$          & $4.73 \pm 0.95$ \givespace
      \hline
      $P_{\rm{c}}$  & days      & $3.69686\,^{+\,0.00036}_{-\,0.00026}$ & $3.70 \pm 0.02$ \givespace
      $K_{\rm{c}}$  & $\ms$     & $5.55\,^{+\,0.34}_{-\,0.31}$          & $6.01 \pm 0.47$ \givespace
      $e_{\rm{c}}$  &           & $0.026\,^{+\,0.033}_{-\,0.017}$       & $0.12 \pm 0.06$ \givespace
      $m_{\rm{c}}$  & $M_\oplus$  & $12.62\,^{+\,0.77}_{-\,0.72}$         & $13.56 \pm 1.08$\givespace
      \hline
      $\eta_3$ \tablefootmark{~(b)} & days & $22.50\,^{+\,10.56}_{-\,6.19}$ & $23.81 \pm 0.03$  \givespace
      \end{tabular}
      \tablefoot{The following notes apply to the estimates from \citetalias{Haywood2014} :\\
      \tablefoottext{a}{A Gaussian prior was used, centred at 0.85359165 days and with a standard deviation of $5.6\!\times\!10^{-7}$ days.} 
      \tablefoottext{b}{The estimate for the rotation period was derived from the CoRoT lightcurve.} }
      \end{table}

\section{Discussion and Conclusions} 
\label{sec:discussion}

  Building on the work of \citet{Brewer2015}, we have developed a simple method to estimate the number of orbiting planets around active stars, using only RV measurements.
  We applied this method to HARPS observations of \corot and confidently detect \corot[b] and \corot[c], while finding weaker evidence for two additional signals.
  In this framework, there is no need to use photometric observations, information from transit detections, or auxiliary activity indicators.

  The posterior distribution for $N_p$ shows evidence for the presence of extra signals, even if they do not meet our detection criteria.
  We note that the effects of considering a uniform prior for $N_p$ (thus giving considerable prior weight to large numbers of planets) and a hierarchical prior for the semi-amplitudes (which changes how the Occam’s razor penalty is taken into account) can be important and will be studied in the future.

  Our choices for the likelihood, covariance function, and priors provide the model with a robustness against possible outliers which is similar to most analyses of RV data.
  If we had reason to suspect the presence of outliers, it would be straightforward to extend our model, e.g. by scaling each error bar with a common value or with individual values whose common prior is defined hierarchically.

  Stepping back to appreciate our results, we find that there is a lot of information contained in the RVs, both about planetary and (arguably more importantly) activity signals, and that this information can be recovered.
  The GP provides a flexible and accommodating model for activity-induced signals, allowing us to infer the planetary masses and orbital parameters with more realistic uncertainties.

  In the analysis of \citetalias{Haywood2014}, the authors modelled the out-of-transit photometry using a GP (as an interpolator) and applied the \emph{FF'} method of \cite{Aigrain2012} to obtain the RV signal that is due to activity.
  By comparing the evidence of models containing this activity signal plus zero, one, two, or three planets, they asserted that the two-planet model was the most probable.
  To model all the quasi-periodic signals in the data, \citetalias{Haywood2014} included a GP (with the covariance properties of the CoRoT light curve) as part of the RV model. This was justified because ``the \emph{FF'} method is likely to provide an incomplete representation of activity-induced RV variations'' \citepalias{Haywood2014}.
  We also note that these authors use a very strong Gaussian prior for $P_{\rm b}$ (and the time of periastron), which was nevertheless completely justified by the transit observations.

  Our analysis rests on much fewer assumptions -- only the RV measurements were used, the rotation period of the star was allowed to vary, the prior for the orbital periods is much less stringent -- but we are still able to recover the same number of confidently detected planets and reliable estimates of the orbital parameters and the stellar rotation period (see Table \ref{tab:parameters}).
  This shows that all this information is contained in the RV measurements and can be recovered with the flexibility of the GP, if we better account for the uncertainty associated with it.

  We should finally note two important properties of the \corot system, which made it ideal for this analysis. 
  First, the amplitudes of the planet signals are much higher than the mean error bar of the HARPS observations, regardless of the stellar activity contamination. 
  Second, the time sampling of the observations is almost ideal for the detection of short-period planets, and is very difficult to obtain as part of a typical RV survey.
  We highlight here the importance of further tests, using other well-studied datasets of active host stars as well as simulated datasets, for assessing the limits of applicability of our method.

  Nevertheless, the importance of this work is not on the specific application to \corot, but instead on providing a simple and fast method to infer the number of planetary signals in the presence of stellar activity.
  With small modifications, this method can be used to search for planets around stars with different activity levels.


\begin{acknowledgements}
The work presented here grew directly from a collaboration that started at the Astro Hack Week 2015.
We thank Andrew Collier Cameron for insightful discussions.
We acknowledge the excellent open-source Python packages made available to the community (in particular \href{http://daft-pgm.org/}{DAFT} and \href{http://dan.iel.fm/george}{George}).
This work was supported by Funda\c{c}\~ao para a Ci\^encia e a Tecnologia (FCT) through the research grant UID/FIS/04434/2013 and the grant PTDC/FIS-AST/1526/2014.
JPF acknowledges support from FCT through the grant reference SFRH/BD/93848/2013.
RDH gratefully acknowledges a grant from the John Templeton Foundation. The opinions expressed in this publication are those of the authors and do not necessarily reflect the views of the John Templeton Foundation.  
BJB is supported by a Fast Start grant from the Royal Society of New Zealand.
PF and NCS acknowledge support by FCT through Investigador FCT contracts (IF/01037/2013 and IF/00169/2012, respectively), and POPH/FSE (EC) by FEDER funding through the program Programa Operacional de Factores de Competitividade - COMPETE. PF further acknowledges support from FCT in the form of an exploratory project of reference IF/01037/2013CP1191/CT0001.
MO acknowledges research funding from the Deutsche Forschungsgemeinschft (DFG , German Research Foundation) - OS 508/1-1.
AS is supported by the European Union under a Marie Curie Intra-European Fellowship for Career Development with reference FP7-PEOPLE-2013-IEF, number 627202.
\end{acknowledgements}

\bibliographystyle{aa}
\bibliography{/home/joao/phd/bib/zotero_library}

\begin{thebibliography}{37}
\expandafter\ifx\csname natexlab\endcsname\relax\def\natexlab#1{#1}\fi

\bibitem[{{Aigrain} {et~al.}(2012){Aigrain}, {Pont}, \& {Zucker}}]{Aigrain2012}
{Aigrain}, S., {Pont}, F., \& {Zucker}, S. 2012, Mon. Not. R. Astron. Soc.,
  419, 3147

\bibitem[{{Baranne} {et~al.}(1996){Baranne}, {Queloz}, {Mayor}, {Adrianzyk},
  {Knispel}, {Kohler}, {Lacroix}, {Meunier}, {Rimbaud}, \& {Vin}}]{Baranne1996}
{Baranne}, A., {Queloz}, D., {Mayor}, M., {et~al.} 1996, \aaps, 119, 373

\bibitem[{{Barros} {et~al.}(2014){Barros}, {Almenara}, {Deleuil}, {Diaz},
  {Csizmadia}, {Cabrera}, {Chaintreuil}, {Collier Cameron}, {Hatzes},
  {Haywood}, {Lanza}, {Aigrain}, {Alonso}, {Bord{\'e}}, {Bouchy}, {Deeg},
  {Erikson}, {Fridlund}, {Grziwa}, {Gandolfi}, {Guillot}, {Guenther}, {Leger},
  {Moutou}, {Ollivier}, {Pasternacki}, {P{\"a}tzold}, {Rauer}, {Rouan},
  {Santerne}, {Schneider}, \& {Wuchterl}}]{Barros2014}
{Barros}, S. C.~C., {Almenara}, J.~M., {Deleuil}, M., {et~al.} 2014, Astron.
  Astrophys., 569, A74

\bibitem[{{Boisse} {et~al.}(2011){Boisse}, {Bouchy}, {H{\'e}brard}, {Bonfils},
  {Santos}, \& {Vauclair}}]{Boisse2011}
{Boisse}, I., {Bouchy}, F., {H{\'e}brard}, G., {et~al.} 2011, Astron.
  Astrophys., 528, A4

\bibitem[{{Brewer}(2014)}]{Brewer2014}
{Brewer}, B.~J. 2014, arXiv:1411.3921

\bibitem[{{Brewer} \& {Donovan}(2015)}]{Brewer2015}
{Brewer}, B.~J. \& {Donovan}, C.~P. 2015, Mon. Not. R. Astron. Soc., 448, 3206

\bibitem[{{Brewer} {et~al.}(2011){Brewer}, {P{\'a}rtay}, \&
  {Cs{\'a}nyi}}]{Brewer2011}
{Brewer}, B.~J., {P{\'a}rtay}, L.~B., \& {Cs{\'a}nyi}, G. 2011, Stat. Comput.,
  21, 649

\bibitem[{{Cover} \& {Thomas}(2006)}]{Cover2006}
{Cover}, T.~M. \& {Thomas}, J.~A. 2006, Elements of information theory, 2nd
  edn. (Hoboken, N.J: {Wiley-Interscience})

\bibitem[{{Dumusque} {et~al.}(2011{\natexlab{a}}){Dumusque}, {Santos}, {Udry},
  {Lovis}, \& {Bonfils}}]{Dumusque2011a}
{Dumusque}, X., {Santos}, N.~C., {Udry}, S., {Lovis}, C., \& {Bonfils}, X.
  2011{\natexlab{a}}, Astron. Astrophys., 527, A82

\bibitem[{{Dumusque} {et~al.}(2011{\natexlab{b}}){Dumusque}, {Udry}, {Lovis},
  {Santos}, \& {Monteiro}}]{Dumusque2011b}
{Dumusque}, X., {Udry}, S., {Lovis}, C., {Santos}, N.~C., \& {Monteiro}, M. J.
  P. F.~G. 2011{\natexlab{b}}, Astron. Astrophys., 525, A140

\bibitem[{{Feroz} {et~al.}(2011){Feroz}, {Balan}, \& {Hobson}}]{Feroz2011}
{Feroz}, F., {Balan}, S.~T., \& {Hobson}, M.~P. 2011, Mon. Not. R. Astron.
  Soc., 415, 3462

\bibitem[{{Ferraz-Mello} {et~al.}(2011){Ferraz-Mello}, {Tadeu dos Santos},
  {Beaug{\'e}}, {Michtchenko}, \& {Rodr{\'\i}guez}}]{Ferraz-Mello2011}
{Ferraz-Mello}, S., {Tadeu dos Santos}, M., {Beaug{\'e}}, C., {Michtchenko},
  T.~A., \& {Rodr{\'\i}guez}, A. 2011, Astron. Astrophys., 531, A161

\bibitem[{{Figueira} {et~al.}(2010){Figueira}, {Marmier}, {Bonfils}, {di
  Folco}, {Udry}, {Santos}, {Lovis}, {M{\'e}gevand}, {Melo}, {Pepe}, {Queloz},
  {S{\'e}gransan}, {Triaud}, \& {Viana Almeida}}]{Figueira2010}
{Figueira}, P., {Marmier}, M., {Bonfils}, X., {et~al.} 2010, Astron.
  Astrophys., 513, L8

\bibitem[{{Figueira} {et~al.}(2013){Figueira}, {Santos}, {Pepe}, {Lovis}, \&
  {Nardetto}}]{Figueira2013}
{Figueira}, P., {Santos}, N.~C., {Pepe}, F., {Lovis}, C., \& {Nardetto}, N.
  2013, Astron. Astrophys., 557, A93

\bibitem[{{Grunblatt} {et~al.}(2015){Grunblatt}, {Howard}, \&
  {Haywood}}]{Grunblatt2015}
{Grunblatt}, S.~K., {Howard}, A.~W., \& {Haywood}, R.~D. 2015, Astrophys. J.,
  808, 127

\bibitem[{{Hatzes} {et~al.}(2010){Hatzes}, {Dvorak}, {Wuchterl}, {Guterman},
  {Hartmann}, {Fridlund}, {Gandolfi}, {Guenther}, \&
  {P{\"a}tzold}}]{Hatzes2010}
{Hatzes}, A.~P., {Dvorak}, R., {Wuchterl}, G., {et~al.} 2010, Astron.
  Astrophys., 520, A93

\bibitem[{{Hatzes} {et~al.}(2011){Hatzes}, {Fridlund}, {Nachmani}, {Mazeh},
  {Valencia}, {H{\'e}brard}, {Carone}, {P{\"a}tzold}, {Udry}, {Bouchy},
  {Deleuil}, {Moutou}, {Barge}, {Bord{\'e}}, {Deeg}, {Tingley}, {Dvorak},
  {Gandolfi}, {Ferraz-Mello}, {Wuchterl}, {Guenther}, {Guillot}, {Rauer},
  {Erikson}, {Cabrera}, {Csizmadia}, {L{\'e}ger}, {Lammer}, {Weingrill},
  {Queloz}, {Alonso}, {Rouan}, \& {Schneider}}]{Hatzes2011}
{Hatzes}, A.~P., {Fridlund}, M., {Nachmani}, G., {et~al.} 2011, Astrophys. J.,
  743, 75

\bibitem[{{Haywood} {et~al.}(2014){Haywood}, {Collier Cameron}, {Queloz},
  {Barros}, {Deleuil}, {Fares}, {Gillon}, {Lanza}, {Lovis}, {Moutou}, {Pepe},
  {Pollacco}, {Santerne}, {S{\'e}gransan}, \& {Unruh}}]{Haywood2014}
{Haywood}, R.~D., {Collier Cameron}, A., {Queloz}, D., {et~al.} 2014, Mon. Not.
  R. Astron. Soc., 443, 2517

\bibitem[{{Jeffreys}(1998)}]{Jeffreys1998}
{Jeffreys}, H. 1998, Theory of probability, 3rd edn. (Oxford: {Oxford
  University Press})

\bibitem[{{Kass} \& {Raftery}(1995)}]{Kass1995}
{Kass}, R.~E. \& {Raftery}, A.~E. 1995, J. Am. Stat. Assoc., 90, 773

\bibitem[{{Kipping}(2013)}]{Kipping2013}
{Kipping}, D.~M. 2013, Mon. Not. R. Astron. Soc. Lett., 434, L51

\bibitem[{{Lanza} {et~al.}(2010){Lanza}, {Bonomo}, {Moutou}, {Pagano},
  {Messina}, {Leto}, {Cutispoto}, {Aigrain}, {Alonso}, {Barge}, {Deleuil},
  {Auvergne}, {Baglin}, \& {Collier Cameron}}]{Lanza2010}
{Lanza}, A.~F., {Bonomo}, A.~S., {Moutou}, C., {et~al.} 2010, Astron.
  Astrophys., 520, A53

\bibitem[{{L{\'e}ger} {et~al.}(2009){L{\'e}ger}, {Rouan}, {Schneider}, {Barge},
  {Fridlund}, {Samuel}, {Ollivier}, {Guenther}, {Deleuil}, {Deeg}, {Auvergne},
  {Alonso}, {Aigrain}, {Alapini}, {Almenara}, {Baglin}, {Barbieri}, {Bruntt},
  {Bord{\'e}}, {Bouchy}, {Cabrera}, {Catala}, {Carone}, {Carpano}, {Csizmadia},
  {Dvorak}, {Erikson}, {Ferraz-Mello}, {Foing}, {Fressin}, {Gandolfi},
  {Gillon}, {Gondoin}, {Grasset}, {Guillot}, {Hatzes}, {H{\'e}brard}, {Jorda},
  {Lammer}, {Llebaria}, {Loeillet}, {Mayor}, {Mazeh}, {Moutou}, {P{\"a}tzold},
  {Pont}, {Queloz}, {Rauer}, {Renner}, {Samadi}, {Shporer}, {Sotin}, {Tingley},
  {Wuchterl}, {Adda}, {Agogu}, {Appourchaux}, {Ballans}, {Baron}, {Beaufort},
  {Bellenger}, {Berlin}, {Bernardi}, {Blouin}, {Baudin}, {Bodin}, {Boisnard},
  {Boit}, {Bonneau}, {Borzeix}, {Briet}, {Buey}, {Butler}, {Cailleau},
  {Cautain}, {Chabaud}, {Chaintreuil}, {Chiavassa}, {Costes}, {Cuna Parrho},
  {de Oliveira Fialho}, {Decaudin}, {Defise}, {Djalal}, {Epstein}, {Exil},
  {Faur{\'e}}, {Fenouillet}, {Gaboriaud}, {Gallic}, {Gamet}, {Gavalda},
  {Grolleau}, {Gruneisen}, {Gueguen}, {Guis}, {Guivarc'h}, {Guterman},
  {Hallouard}, {Hasiba}, {Heuripeau}, {Huntzinger}, {Hustaix}, {Imad},
  {Imbert}, {Johlander}, {Jouret}, {Journoud}, {Karioty}, {Kerjean},
  {Lafaille}, {Lafond}, {Lam-Trong}, {Landiech}, {Lapeyrere}, {Larqu{\'e}},
  {Laudet}, {Lautier}, {Lecann}, {Lefevre}, {Leruyet}, {Levacher}, {Magnan},
  {Mazy}, {Mertens}, {Mesnager}, {Meunier}, {Michel}, {Monjoin}, {Naudet},
  {Nguyen-Kim}, {Orcesi}, {Ottacher}, {Perez}, {Peter}, {Plasson}, {Plesseria},
  {Pontet}, {Pradines}, {Quentin}, {Reynaud}, {Rolland}, {Rollenhagen},
  {Romagnan}, {Russ}, {Schmidt}, {Schwartz}, {Sebbag}, {Sedes}, {Smit},
  {Steller}, {Sunter}, {Surace}, {Tello}, {Tiph{\`e}ne}, {Toulouse}, {Ulmer},
  {Vandermarcq}, {Vergnault}, {Vuillemin}, \& {Zanatta}}]{Leger2009}
{L{\'e}ger}, A., {Rouan}, D., {Schneider}, J., {et~al.} 2009, Astron.
  Astrophys., 506, 287

\bibitem[{{Mayor} {et~al.}(2003){Mayor}, {Pepe}, {Queloz}, {Bouchy},
  {Rupprecht}, {Lo Curto}, {Avila}, {Benz}, {Bertaux}, {Bonfils}, {Dall},
  {Dekker}, {Delabre}, {Eckert}, {Fleury}, {Gilliotte}, {Gojak}, {Guzman},
  {Kohler}, {Lizon}, {Longinotti}, {Lovis}, {Megevand}, {Pasquini}, {Reyes},
  {Sivan}, {Sosnowska}, {Soto}, {Udry}, {van Kesteren}, {Weber}, \&
  {Weilenmann}}]{Mayor2003}
{Mayor}, M., {Pepe}, F., {Queloz}, D., {et~al.} 2003, The Messenger, 114, 20

\bibitem[{{Noyes} {et~al.}(1984){Noyes}, {Hartmann}, {Baliunas}, {Duncan}, \&
  {Vaughan}}]{Noyes1984}
{Noyes}, R.~W., {Hartmann}, L.~W., {Baliunas}, S.~L., {Duncan}, D.~K., \&
  {Vaughan}, A.~H. 1984, Astrophys. J., 279, 763

\bibitem[{{Pepe} {et~al.}(2002){Pepe}, {Mayor}, {Galland}, {Naef}, {Queloz},
  {Santos}, {Udry}, \& {Burnet}}]{Pepe2002}
{Pepe}, F., {Mayor}, M., {Galland}, F., {et~al.} 2002, Astron. Astrophys., 388,
  632

\bibitem[{{Pont} {et~al.}(2011){Pont}, {Aigrain}, \& {Zucker}}]{Pont2011a}
{Pont}, F., {Aigrain}, S., \& {Zucker}, S. 2011, Mon. Not. R. Astron. Soc.,
  411, 1953

\bibitem[{{Queloz} {et~al.}(2009){Queloz}, {Bouchy}, {Moutou}, {Hatzes},
  {H{\'e}brard}, {Alonso}, {Auvergne}, {Baglin}, {Barbieri}, {Barge}, {Benz},
  {Bord{\'e}}, {Deeg}, {Deleuil}, {Dvorak}, {Erikson}, {Ferraz Mello},
  {Fridlund}, {Gandolfi}, {Gillon}, {Guenther}, {Guillot}, {Jorda}, {Hartmann},
  {Lammer}, {L{\'e}ger}, {Llebaria}, {Lovis}, {Magain}, {Mayor}, {Mazeh},
  {Ollivier}, {P{\"a}tzold}, {Pepe}, {Rauer}, {Rouan}, {Schneider},
  {Segransan}, {Udry}, \& {Wuchterl}}]{Queloz2009}
{Queloz}, D., {Bouchy}, F., {Moutou}, C., {et~al.} 2009, Astron. Astrophys.,
  506, 303

\bibitem[{{Rajpaul} {et~al.}(2015){Rajpaul}, {Aigrain}, {Osborne}, {Reece}, \&
  {Roberts}}]{Rajpaul2015}
{Rajpaul}, V., {Aigrain}, S., {Osborne}, M.~A., {Reece}, S., \& {Roberts}, S.
  2015, Mon. Not. R. Astron. Soc., 452, 2269

\bibitem[{{Rasmussen} \& {Williams}(2006)}]{Rasmussen2006}
{Rasmussen}, C.~E. \& {Williams}, C. K.~I. 2006, Gaussian processes for machine
  learning (Cambridge: {MIT Press})

\bibitem[{{Roberts} {et~al.}(2012){Roberts}, {Osborne}, {Ebden}, {Reece},
  {Gibson}, \& {Aigrain}}]{Roberts2012}
{Roberts}, S., {Osborne}, M., {Ebden}, M., {et~al.} 2012, Philos. Trans. R.
  Soc. Lond. Math. Phys. Eng. Sci., 371

\bibitem[{{Robertson} {et~al.}(2015){Robertson}, {Roy}, \&
  {Mahadevan}}]{Robertson2015a}
{Robertson}, P., {Roy}, A., \& {Mahadevan}, S. 2015, Astrophys. J., 805, L22

\bibitem[{{Saar} \& {Donahue}(1997)}]{Saar1997}
{Saar}, S.~H. \& {Donahue}, R.~A. 1997, Astrophys. J., 485, 319

\bibitem[{{Santos} {et~al.}(2010){Santos}, {Gomes da Silva}, {Lovis}, \&
  {Melo}}]{Santos2010a}
{Santos}, N.~C., {Gomes da Silva}, J., {Lovis}, C., \& {Melo}, C. 2010, Astron.
  Astrophys., 511, A54

\bibitem[{{Santos} {et~al.}(2014){Santos}, {Mortier}, {Faria}, {Dumusque},
  {Adibekyan}, {Delgado-Mena}, {Figueira}, {Benamati}, {Boisse}, {Cunha},
  {Gomes da Silva}, {Lo Curto}, {Lovis}, {Martins}, {Mayor}, {Melo}, {Oshagh},
  {Pepe}, {Queloz}, {Santerne}, {S{\'e}gransan}, {Sozzetti}, {Sousa}, \&
  {Udry}}]{Santos2014}
{Santos}, N.~C., {Mortier}, A., {Faria}, J.~P., {et~al.} 2014, Astron.
  Astrophys., 566, A35

\bibitem[{{Tuomi}(2011)}]{Tuomi2011}
{Tuomi}, M. 2011, Astron. Astrophys., 528, L5

\bibitem[{{Tuomi} {et~al.}(2014){Tuomi}, {Anglada-Escude}, {Jenkins}, \&
  {Jones}}]{Tuomi2014}
{Tuomi}, M., {Anglada-Escude}, G., {Jenkins}, J.~S., \& {Jones}, H. R.~A. 2014,
  arXiv:1405.2016

\end{thebibliography}

\end{document}